\documentclass[a4paper,12pt, oneside]{article}
\usepackage{graphicx}
\usepackage{dcolumn}
\usepackage{bm}

\begin{document}
\baselineskip 24pt

\begin{center}
{\Large Depth profile high-energy spectroscopic study of Mn-doped GaN
prepared by thermal diffusion}
\vspace{1cm}

{J. I. Hwang, Y. Osafune, M. Kobayashi, K. Ebata, Y. Ooki,\\ Y. Ishida and A. Fujimori}\\
\it{Department of Complexity Science and Engineering, Department of Physics, University of Tokyo, Kashiwanoha, Kashiwa-shi, Chiba, 277-8561, Japan}
\vspace{0.5cm}

\rm{Y. Takeda, T. Okane and Y. Saitoh}\\
\it{Synchrotron Radiation Research Unit, JAEA, SPring-8, Sayo-gun, Hyogo 679-5148, Japan}
\vspace{0.5cm}

\rm{K. Kobayashi}\\
\it{Synchrotron Radiation Research Unit, JAEA, SPring-8, Sayo-gun, Hyogo 679-5148, Japan and JASRI, SPring-8, Sayo-gun, Hyogo 679-5198, Japan}
\vspace{0.5cm}

\rm{A. Tanaka}\\
\it{Department of Quantum Matter, ADSM, Hiroshima University, Kagamiyama, Higashi-Hiroshima-shi, Hiroshima 739-8530, Japan}
\vspace{0.5cm}

\end{center}


\newpage
\begin{center}
\section*{Abstract}
\end{center}
We have performed an $in$-$situ$ depth profile study of Mn-doped GaN prepared by a low temperature thermal diffusion method using photoemission and x-ray absorption spectroscopy. 
It was revealed from the core-level photoemission measurements that Mn ions are diffused into a deep ($\sim$ 70 \AA) region of the GaN substrates and that the line shapes of Mn 3$d$ partial density of states obtained by  resonant photoemission measurements was close to that of Ga$_{1-x}$Mn$_x$N thin films grown by molecular-beam epitaxy. From x-ray absorption spectroscopy (XAS) and x-ray magnetic circular dichroism (XMCD) measurements at the Mn $L$-edge, it was revealed that the doped Mn ions were in the divalent Mn$^{2+}$ state and primarily paramagnetic. In magnetization measurements, 
weak hysteresis was detected in samples prepared using $p$-type GaN substrates while samples using $n$-type GaN substrates showed only paramagnetism.


\newpage

\section{\label{sec:intro}Introduction}
%
%
%
%
After the discovery of carrier-induced ferromagnetism in Ga$_{1-x}$Mn$_x$As and In$_{1-x}$Mn$_x$As, diluted magnetic semiconductors (DMSs) have attracted much interest both in physics of spin-related transport and applications for spintronics. However, because the Curie temperatures ($T_C$'s) of Ga$_{1-x}$Mn$_x$As and In$_{1-x}$Mn$_x$As are far below room temperature, one of the most urgent issues in this research filed has been to synthesize DMSs which exhibit ferromagnetism at room temperature. Recent theoretical studies have predicted that GaN and ZnO doped with transition-metal ions may show ferromagnetism at room temperature \cite{5Dietl, 5Sato, 5Mahadevan}. Nitride-based DMSs such as Mn-doped GaN have therefore been extensively studied. 
So far, Mn doping into GaN has been achieved mainly through molecular-beam epitaxy (MBE) \cite{5Kuwabara, 5Zajac, 5Sonoda, 5HashimotoMn}.
Recently, Sonoda $et$ $al.$ \cite{5SonodaAPL} and Sarigiannidou $et$ $al.$ \cite{5SariPRB} have reported that high quality Ga$_{1-x}$Mn$_x$N thin films grown by MBE on wurtzite GaN substrates showed low temperature ferromagnetism and that the valence of the Mn ion was 3+. Besides, Sonoda $et$ $al.$ \cite{5SonodaAPL} demonstrated that the Mn valence can be changed from 3+ to 2+ by hydrogenation and the ferromagnetism is enhanced in such a system in which Mn$^{2+}$ and Mn$^{3+}$ coexist with $p$-type carriers, indicating the possibility of high temperature ferromagnetism in this system.

%
%
%
%
Mn doping into GaN has also been achieved by low-temperature thermal diffusion methods \cite{5Reed, 5Cai}, in which Mn metal was deposited on the GaN surfaces and the samples were subsequently annealed above 250 $^\circ$C. It has been reported that Mn atoms were detected in deep regions of the samples regardless of annealing temperature and that the samples thus prepared indicated ferromagnetic behavior at room temperature. Thermal diffusion phenomena have also been widely studied in transition-metal-ion-doped semiconductors such as V-, Cr-, and Mn-doped GaAs \cite{5Bchetnia, 5Kutt, 5Hilton} because thermal diffusion process has played an important roles in semiconductor device fabrication. Thermal diffusion process is expected to be important for magnetism of other DMSs such as Mn doped GaAs \cite{5Chiba, 5Kirby} and Mn doped CdGeP$_2$ \cite{5CdGeP} and ZnGeP$_2$ \cite{5ZnGeP}, too. Ishida $et\ al.$ \cite{5Ishida} revealed performing the depth profiling studies of ZnGeP$_2$:Mn prepared using the thermal diffusion method that the Mn$^{2+}$ substitution for Zn$^{2+}$ was induced and that the ferromagnetism in ZnGeP$_2$:Mn is caused by the dilute-Mn ions in the deep region. Thus, the investigation of the thermal diffusion depth profile and the electronic structure in various depth regions at each stage of thermal diffusion will provide us with important information to understand the ferromagnetism of those DMSs. So far, detail investigations of Mn-doped GaN by the thermal diffusion method have been lacking compared with the investigation of MBE grown Mn-doped GaN. In the present study, we have applied various high-energy spectroscopic techniques to study the depth profile studies of Mn-doped GaN {\it in-situ} prepared by the low temperature thermal diffusion method.

%
%
%
%
\section{\label{sec:exp}Experimental}
GaN substrates used in this study were 2$\mu$m-thick $p$- and $n$-type GaN epitaxial layers grown by metal-organic chemical vapor deposition (MOCVD) on a 1$\mu$m-thick undoped GaN layer, which had been grown on a 30 nm-thick buffer layer grown on a sapphire (0001) substrate (POWDEC Co., Ltd.). High-quality Mn metal (purity 99.9995 at \%) was used as the evaporant. Before the Mn deposition, surface contamination and oxidized layer of the substrate were removed by N$\rm_2$-ion sputtering followed by 500 $^{\circ}$C annealing. The surface cleanliness was checked by O 1$s$ and C 1$s$ core-level photoemission spectra. Then 3 nm-thick Mn was deposited on the GaN surface at room temperature. After the Mn deposition, the samples were post-annealed at 500 $^{\circ}$C for 6 hours. Depth profile analysis was achieved by repeated cycles of Ar-ion sputtering and photoemission measurement. The incident angle of the ion beam was set to 75$^{\circ}$ from the surface normal. The sputter-etching rate was 0.02 nm\slash min. As a reference, the depth profile analysis of GaN without Mn was also performed. 

%
%
%
%
The {\it in-situ} depth profile experiments were performed using core-level x-ray photoemission spectroscopy (XPS) using Mg$K\alpha$ source and resonant photoemission spectroscopy (RPES) using synchrotron radiation. In the XPS measurements, photoelectrons were collected using a VSW multi-channel analyzer. The $K\alpha_{2}$, $K\alpha_{3,4}$ and $K\alpha_{5,6}$ ghosts have been numerically subtracted. The RPES measurements using synchrotron radiation were performed at BL-18A of Photon Factory (PF), High Energy Accelerator Research Organization (KEK). Photoelectrons were collected using a VG CLAM hemispherical analyzer. The total energy resolution including the monochromator, the electron analyzer and the temperature broadening was $\sim$200 meV for RPES and $\sim$900 meV for XPS. The photoemission spectra were referenced to the Fermi edge of a metal in electrical contact with the sample. XAS and XMCD measurements were performed at the soft x-ray beam line BL-23SU of SPring-8. Absorption spectra were measured by the total electron yield method with the energy resolution $E$/$\Delta$$E$ better than 10000. All the measurements were made in an ultra-high vacuum below 1$\times$10$^{-10}$ Torr at room temperature.

%
%
%
%
\section{\label{sec:depth}Depth profile analysis}
Figure 1 shows the core-level photoemission spectra recorded in the sputter-etching series and provides the depth profile.  Sputtering time $t$ is indicated in the figure. In Fig. 1 (a) and (b), Ga 2$p_{3/2}$ and the N 1$s$ signals were observed for the as-annealed surface ($t$ = 0 min), indicating the outdiffusion of Ga and N atoms to the surface. Near the sample surface ($t$ = 0 - 120 min), Ga 2$p_{3/2}$ and the N 1$s$ spectra consisted of three components as indicated by dashed vertical lines, at the binding energy ($E_B$) of 1116.9 (A), 1117.8 (B) and 1119.0 eV (C) for Ga 2$p_{3/2}$ and 396.1 (C'), 397.3 (B') and 398.0 eV (A') for N 1$s$, while Mn metal was dominant in the Mn 2$p$ spectra as shown in Fig. 1 (c). After 120 min sputtering, the intensities of Ga 2$p_{3/2}$ (B) and N 1$s$ (B') rapidly increased while that of Mn 2$p$ and other components of Ga 2$p_{3/2}$ and the N 1$s$ were rapidly decreased. After 240 min sputtering, Ga 2$p_{3/2}$ (B) and N 1$s$ (B') became dominant and these intensities remained nearly unchanged for further sputtering. These peaks are attributed to GaN because the energy difference between Ga 2$p_{3/2}$ and N 1$s$ in this region ($720.5 \pm 0.2$ eV) agreed with that of a reference GaN sample ($720.8 \pm 0.2$ eV). This indicates that the other components of Ga 2$p_{3/2}$ and N 1$s$ near the surface region (A, C, A' and C') were due to Ga and N atoms in metallic Mn compounds.

%
%
%
%
Figure 2 shows chemical compositions at each step of the sputter-etching series estimated from the intensities of the core-level spectra. The Ga 2$p_{3/2}$ and N 1$s$ intensities have been normalized to those of the reference GaN sample whereas the Mn 2$p$ intensity has been normalized to that of Mn metal. In addition, these intensities have been renormalized so that the total intensities of Ga, N and Mn at each step are equal to unity. The chemical compositions thus deduced indicated a characteristic atomic composition profile as shown in Fig. 2 (a), in good agreement with the results of secondly ion mass spectroscopy (SIMS) studies previously reported for Mn diffusion \cite{5Reed2, 5Cai}. The rapid change of the Mn concentration in the depth profile has been attributed to the relatively small diffusion coefficient of Mn in GaN \cite{5Cai}. The decomposed compositions of Ga 2$p_{3/2}$ and N 1$s$ are shown in Fig. 2 (b) and (c), respectively. It was revealed that Ga 2$p_{3/2}$ (A) and N 1$s$ (A') showed similar composition profiles to Mn 2$p$, implying these components are due to Mn metallic compound. The other components of N 1$s$ (C') and Ga 2$p_{3/2}$ (C), especially C', are mainly distributed near the sample surface.

%
%
%
%
%
%
 With a closer inspection of the composition profile [Fig. 2 (a)], one can see that in the deep region (after 250 min sputtering) the Ga intensity was slightly reduced compared with that of the reference GaN sample whereas weak Mn intensity of $\sim$ 1 \% per cation site existed [see a dashed line in Fig. 2 (a)]. This can be understood that the Mn atoms in the deep region are substituting for the Ga sites of the GaN host. The Mn 2$p$ spectra near the sample surface ($t$ = 0 min) and that in the deep region ($t$ = 350 min) are compared in Fig. 3. The spectra in the deep region were found to be shifted toward higher binding energies compared to the metallic one near the sample surface. The peak position of 641.5 $\pm$ 0.1 eV was close to that of Mn$^{2+}$ in Ga$_{1-x}$Mn$_x$N grown by MBE \cite{5Hwang} (641.4 eV) and in Zn$_{1-x}$Mn$_x$O grown by pulsed-laser deposition\cite{5Mizokawa} (641.0 eV). This indicates that the valence state of Mn in this region was close to Mn$^{2+}$ and different from that of metallic Mn.

%
%
%
%
%
%
Figure 4 shows the Mn 3$p$-3$d$ absorption spectra and the valence-band photoemission spectra of the $p$-GaN:Mn sample in the Mn 3$p$-3$d$ core excitation region. Figure 4 (a) shows the absorption spectra recorded in the total electron yield mode around the Mn 3$p$-3$d$ core-excitation threshold for various depths. With increasing depth, the absorption edge due to metallic Mn at $h\nu = 46.2$ eV was gradually reduced while another edge appeared at $h\nu = 49$ eV . The absorption spectra after 350 min sputtering (shown in the inset) revealed a peak at $h\nu = 50$ eV characteristic of the Mn$^{2+}$ ion and did not show the metallic feature (i.e. the threshold at $h\nu = 46.2$ eV). Figure 4 (b), (c), (d), and (e) show valence-band photoemission spectra in the Mn 3$p$-3$d$ core-excitation region in the sputtering series. In the spectra before 120 min sputtering [Fig. 4 (b) and (c)], a clear Fermi edge was observed with an overlapping Mn $M_{2,3}M_{4,5}M_{4,5}$ Auger emission as indicated by triangles in the figure. After 200 min sputtering, the Fermi edge and the Auger emission disappeared and Mn 3$d$ resonant photoemission, which is caused by the interference between normal photoemission and 3$p$-to-3$d$ transition followed by a 3$p$-3$d$-3$d$ super-Coster-Kr\"onig decay, became dominant. This indicates that the character of the Mn 3$d$ states changed from itinerant to localized one with increasing depth. This is consistent with the energy shift of the Mn 2$p$ core-level photoemission spectra toward higher binding energies (Fig. 3) and the change of the line shape in the Mn 3$p$-3$d$ absorption spectra. From RPES, 
one can obtain the Mn 3$d$ partial density of states (PDOS) by subtracting the off-resonant spectra ($h\nu = 48.5$ eV) from the on-resonance ($h\nu = 50$ eV) one. The Mn 3$d$ PDOS thus obtained is shown in the bottom panels of Fig. 4 (d) and (e). The Mn 3$d$ PDOS indicated a peak at $E_B$ = 5 eV, a shoulder at $E_B$ = 2 eV and a satellite structure at $E_B$ = 10 eV, similar to the Mn 3$d$ PDOS of MBE-grown Ga$_{1-x}$Mn$_x$N \cite{5Hwang}. This indicates that the Mn ions in the deep region have similar electronic environments to those of MBE-grown Ga$_{1-x}$Mn$_x$N.

%
%
%
%
%
%

Summarizing the depth profile study, it has been revealed that the chemical composition and the electronic structure were quite different before and after 140 min sputtering. In the region exposed before 120 min sputtering, Mn was metallic including the Mn-Ga-N compounds. In the region exposed after 200 min sputtering, GaN was dominant and Mn$^{2+}$ ions were diluted in the GaN host.  In this region, dilute Mn ions existed while the Ga concentration was slightly reduced. Further, its electronic structure was similar to that of Ga$_{1-x}$Mn$_x$N grown by MBE. These results suggest that a diluted magnetic semiconductor was formed in the deep region prepared by the thermal diffusion method.

%
%
%
%
%
\section{\label{sec:XAS}X-ray magnetic circular dichroism and magnetic properties}
In order to further investigate the electronic state of Mn ions in the deep region, we have also performed XAS measurements. Figure 5 (a) shows XAS spectra at the Mn $L_{2,3}$ edge of the $p$-GaN:Mn sample after 350 min sputtering compared with atomic multiplet calculations for the $d^{4}$ and $d^{5}$ configurations in a tetrahedral crystal field. The observed spectrum shows multiplet structures typical for localized 3$d$ states, and resembles that of Mn$^{2+}$ in Ga$_{1-x}$Mn$_x$N grown by MBE\cite{5Hwang, 5Edmonds}. Comparing the experimental spectra with the calculations, one can see that the calculation for the $d^{5}$ configuration in the tetrahedral crystal field 10$Dq$ = -0.5 eV well reproduced the experimental spectra. Figure 5 (b) shows the XAS spectrum for right-polarized light $\mu_{+}$ [solid curve in Fig. 5 (b)], left-polarized light $\mu_{-}$ (dashed curve) and the difference $\mu_{+} - \mu_{-}$ (XMCD: middle panel) in the magnetic field of 5 T at 30 K. The XMCD intensity was 16 \% of the XAS intensity corresponding to 34 \% of the full polarization of the Mn$^{2+}$ ions. The XMCD spectra were also well explained by the atomic multiplet calculation assuming the $d^{5}$ ground state with 10$Dq$ = -0.5 eV. In the XMCD spectra, structures appeared at the same energies as XAS indicating that the divalent Mn ions in the tetrahedral crystal field were responsible for the obtained XMCD signal. Figure 5 (c) shows the magnetic field dependence of the XMCD signal and hence the magnetism of the $p$-GaN:Mn sample. No clear change in the overall spectral line shape was observed between the spectra recorded at different magnetic fields as shown in the inset of Fig. 5 (c), indicating that other ferromagnetic phases were not detected in this region. The XMCD intensity showed linear response to the magnetic field, indicating that the divalent Mn ions were mostly paramagnetic. 

%
%
%
%
%
We have also performed magnetization measurements using a superconducting quantum interference device (SQUID). Figure 6 (a) shows temperature dependence of magnetic susceptibility per Mn ion at a high magnetic fields ($H >$ 4 T). The data were fitted to the Curie-Weiss law plus a temperature independent constant, ${\partial M}/{\partial H_{H>4{\rm T}}} = NC/(T-\Theta) + \chi_0$. Here, $N$ is number of the magnetic ions, $g$ is the $g$ factor, $C = (g\mu_B)^2S(S+1)/3k_B$ is the Curie constant and $\theta$ is the Weiss temperature. Assuming $S=5/2$, good fit was obtained with $N = {\rm (} 2.48 \pm 0.01{\rm )} \times 10^{15}$, $\Theta = - 7.7 \pm 2.0$ K and $\chi_0 = {\rm (-}3.36 \pm 0.01{\rm )} \times 10^{16}\mu_{\rm B}/{\rm T}$, indicating that the temperature dependence of the susceptibility is caused by local magnetic moments. The negative constant $\chi_0$ came from the sapphire substrate and the GaN host. The negative Weiss temperature indicates antiferromagnetic interaction between the local magnetic moments. The number $N$ corresponds to averaged concentration $2.5 \times 10^{20} {\rm cm}^{-3}$ ($x \sim 0.6 \%$). This value is smaller than that obtained from the depth profile using XPS ($x \sim 1 \%$) [see Fig. 2 (a)]. This is probably due to the decreasing Mn distribution toward deep region caused by thermal diffusion and hence the higher Mn concentration near the sample surface \cite{5Cai}. Nevertheless, since this value fairly well corresponds to that obtained from depth profile using XPS, we attribute the local magnetic moments to those of Mn ions.
The magnetic susceptibility at 30 K estimated from XMCD using XMCD sum rules \cite{5sum1, 5sum2} is also shown in Fig. 6 (a). The susceptibility per Mn atom obtained from the SQUID measurement was larger than that obtained from XMCD by a factor of $\sim$ 1.6. This discrepancy is probably due to the difference between the average Mn concentration detected by a SQUID and that near the sample surface detected by XMCD as mentioned above. Thus, the paramagnetic susceptibility obtained from XMCD can be explained by the Curie-Weiss law. 

%
%
%
%
The sample prepared using the $p$-type GaN substrate ($t$ = 350 min), however, showed a weak hysteresis up to 300 K, as shown in Fig. 6 (b). Here, the diamagnetic component has been subtracted using $\chi_0$. Although the magnetic field dependence of the XMCD signal in Fig. 5 (b) should in principle show the corresponding behavior, the ferromagnetic signal was too weak to be detected in the present XMCD measurement. Interestingly, hysteresis was not observed in the thermal diffusion sample prepared using the Si-doped $n$-type GaN substrate as shown in Fig. 6 (c). Although it has been reported that intentionally hole-doped Ga$_{1-x}$Mn$_x$N (through Mg co-doing \cite{5Kim}, charge transfer \cite{5Arkun}, etc) shows ferromagnetism while conventional Ga$_{1-x}$Mn$_x$N show paramagnetism, it is not clear at present whether the ferromagnetic component was intrinsic or comes from a tiny amount of segregation undetected in this study. Further systematic studies, particularly, highly sensitive XMCD measurements are necessary to clarify whether the ferromagnetism in the $p$-type GaN:Mn was intrinsic or not.

%
%
%
%
\section{\label{sec:conclusion}Conclusion}
We have performed an $in$-$situ$ depth profile study of Mn-doped GaN prepared by the low-temperature thermal diffusion method. From the depth profile studies using core-level photoemission spectroscopy, it has been revealed that dilute Mn ions were incorporated into GaN host in the deep region while the surface region remained as the metallic layer consisted of the Mn-Ga-N metallic compound. In the deep region, we have observed spectral feature of the localized Mn 3$d$ state similar to that of the MBE-grown Ga$_{1-x}$Mn$_x$N thin film, showing  the formation of the diluted magnetic semiconductor Ga$_{1-x}$Mn$_x$N in the deep region. In this region, the XMCD spectra indicated no significant changes in the line shape down to $H$ = 0.3 T and that the sample was primarily paramagnetic. Nevertheless, SQUID measurements revealed very weak ferromagnetic signals in the sample prepared on the $p$-type substrate.

\newpage
\section*{\label{sec:acknowledge}ACKNOWLEDGEMENTS}
We thank A. Harasawa and T. Okuda for her variable technical support for the experiment at KEK-PF. This work was supported by a Grant-in-Aid for Scientific Research in Priority Area 
``Semiconductor Nano-Spintronics'' (14076209). The experiment at KEK-PF 
was done under the approval of the Photon Factory Program Advisory 
Committee (2004G002), and that at SPring-8 under the approval of the SPring-8 
Proposal Review Committee (2004B0449-NSa-np-Na).  We also thank the Material 
Design and Characterization Laboratory,  Institute for Solid State 
Physics, University of Tokyo, for the use of the SQUID magnetometer.

\newpage
\begin{figure}
\includegraphics[height=5cm]{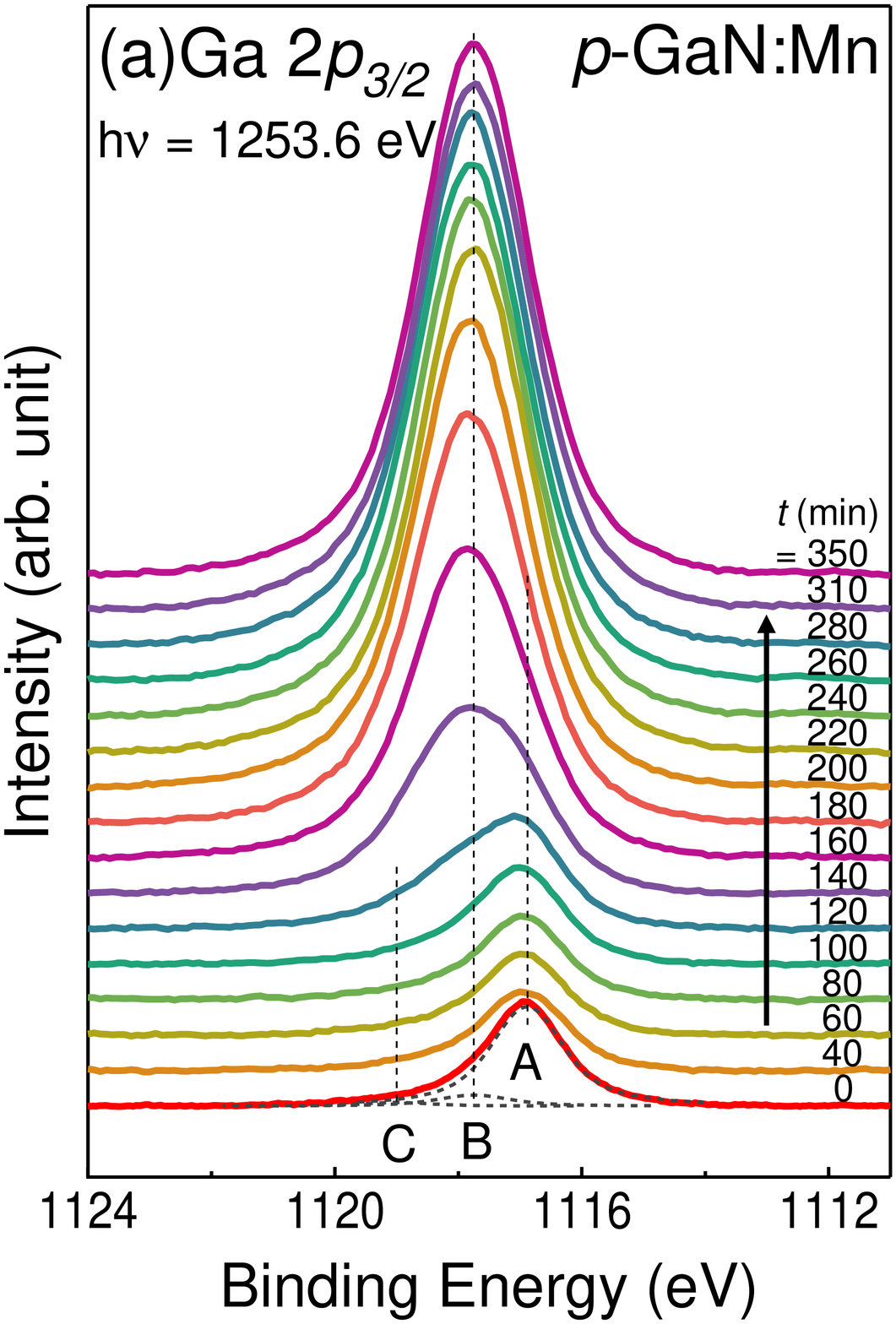}
\includegraphics[height=5cm]{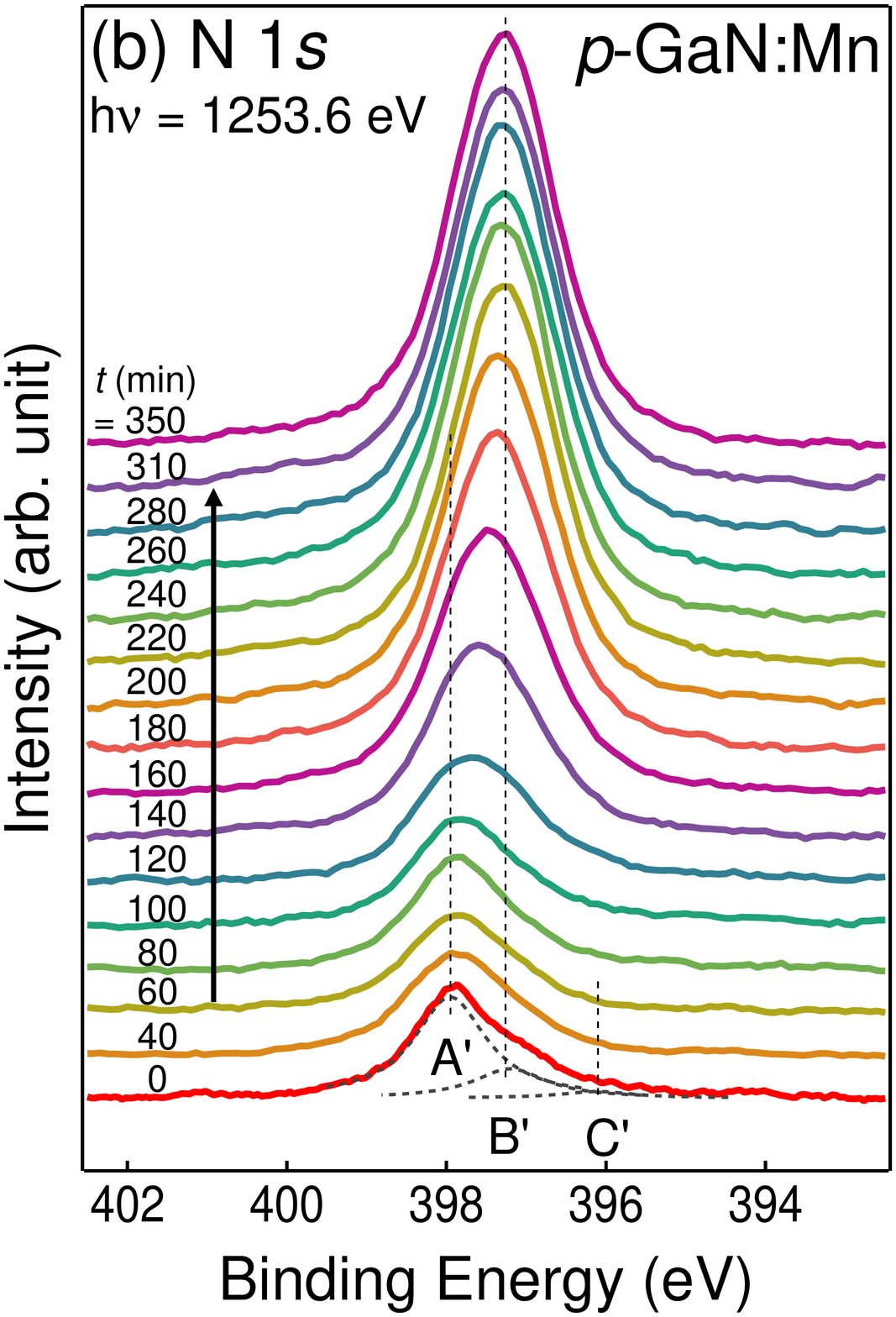}
\includegraphics[height=5cm]{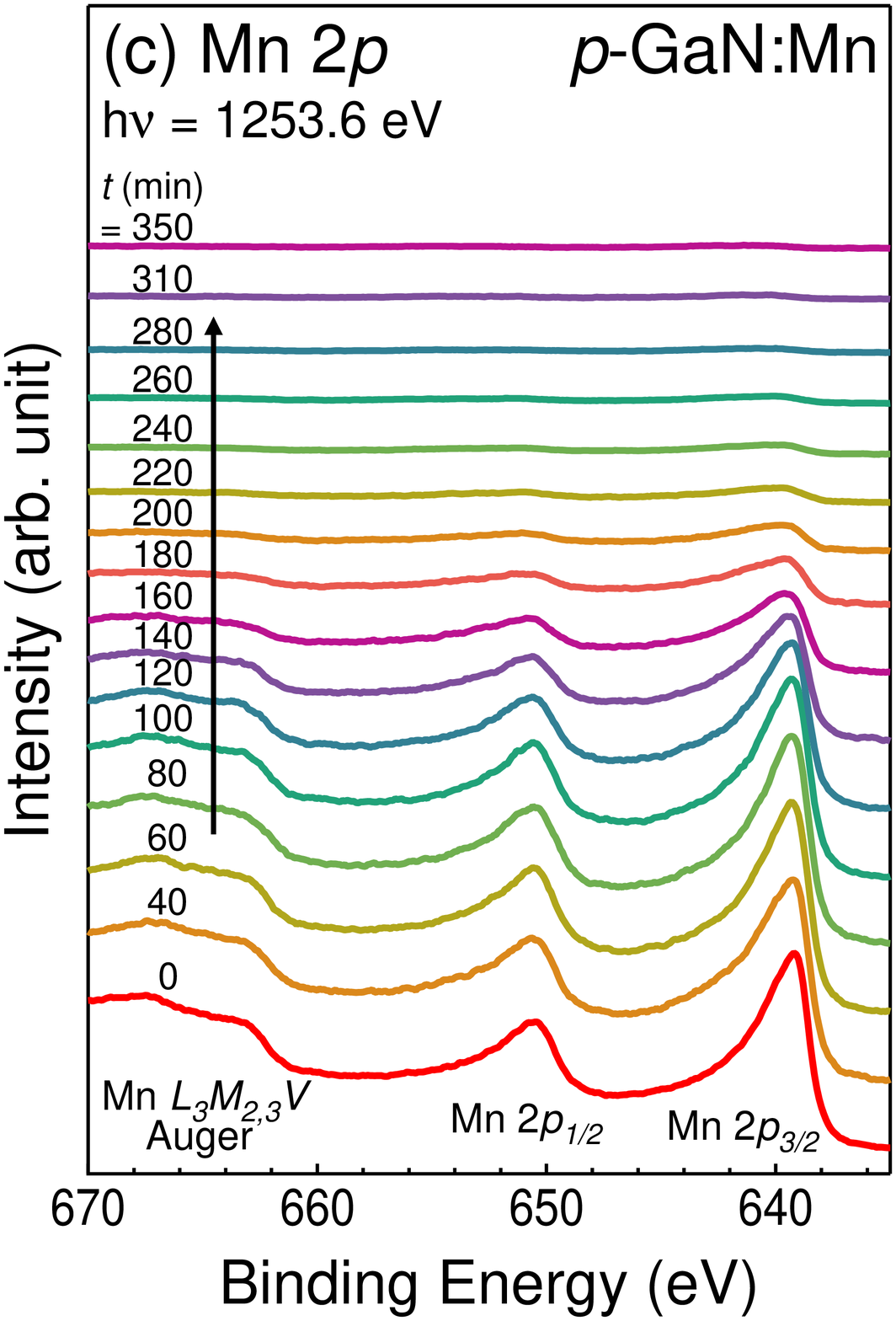}
\label{Int_depth}
\caption{(Color online) Core-level photoemission spectra of $p$-GaN:Mn recorded in the sputter-etching series. Raw data of Ga 2$p_{3/2}$ (a), N 1$s$ (b) and Mn 2$p$ (c) core levels. $t$ = 0 min corresponds to the as-annealed surface. In (a) and (b), the Ga 2$p_{3/2}$ and N 1$s$ core levels have been decomposed to three components. The sputter-etching rate was 0.02 nm\slash min.}
\end{figure}

\vspace{1cm}
\begin{figure}
\includegraphics[height=5cm]{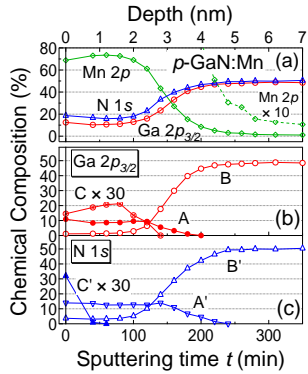}
\label{Int_depth_atom}
\caption{(Color online) Chemical compositions during the sputtering process of $p$-GaN:Mn. (a) Total amount of each atom. (b) Each component of Ga 2$p_{3/2}$ [see Fig. 1 (a)]. (c) Each component of N 1$s$ [see Fig. 1 (a)].}
\end{figure}

\vspace{1cm}
\begin{figure}
\includegraphics[height=5cm]{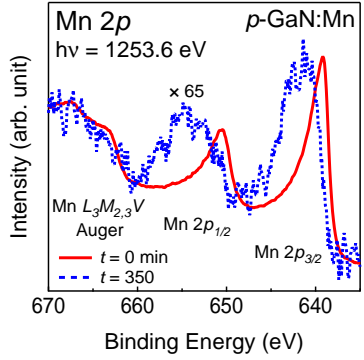}
\label{Mn2p_comp}
\caption{(Color online) Mn 2$p$ core-level photoemission spectra of $p$-GaN:Mn at $t$ = 0 min (solid line) and $t$ = 350 min (dashed line).}
\end{figure}

\vspace{1cm}
\begin{figure}
\includegraphics[height=5cm]{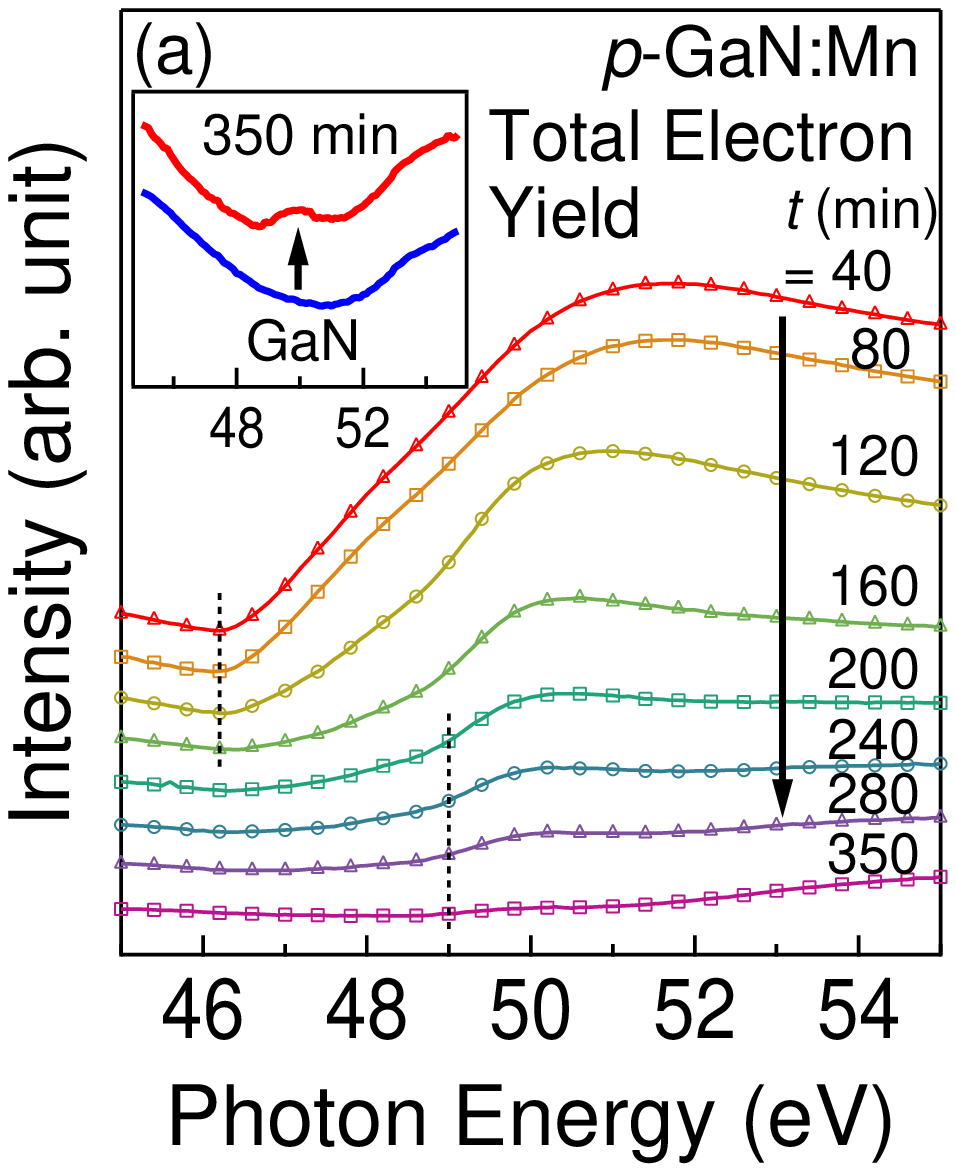}
\includegraphics[height=5cm]{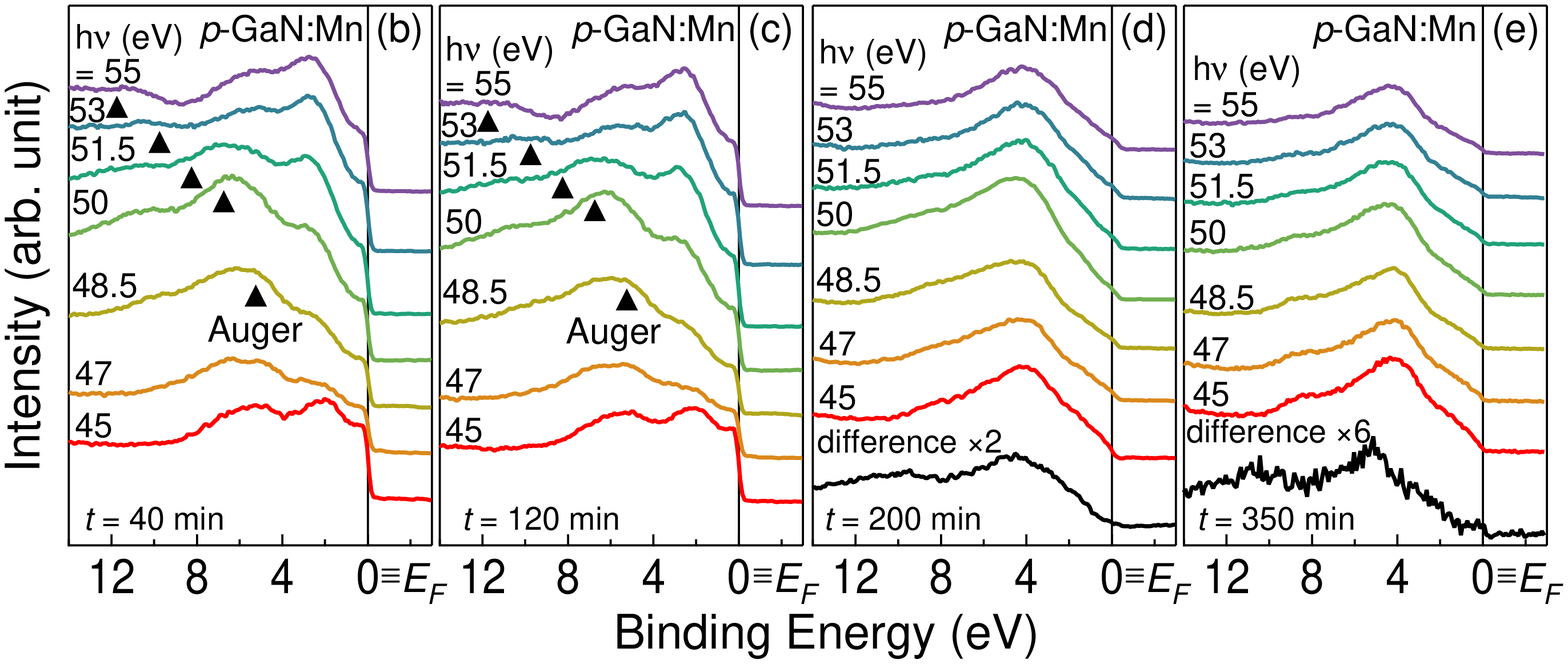}
\label{RPES}
\caption{(Color online) Mn 3$p$-3$d$ absorption spectra and valence-band photoemission spectra of $p$-GaN:Mn for various photon energies in the Mn 3$p$-3$d$ core-excitation region. (a) Mn 3$p$-3$d$ absorption spectra recorded by the total electron yield method for various depths. The inset shows comparison between the spectrum after 350 min sputtering and that of the reference GaN sample. (b) - (e) Valence-band photoemission spectra. Triangles in (b) and (c) denote Mn $M_{2,3}M_{4,5}M_{4,5}$ Auger emission. The difference curves at the bottom of (d) and (e) represent the Mn 3$d$ PDOS. The sputter-etching rate was 0.02 nm\slash min.}
\end{figure}

\vspace{1cm}
\begin{figure}
\includegraphics[height=5cm]{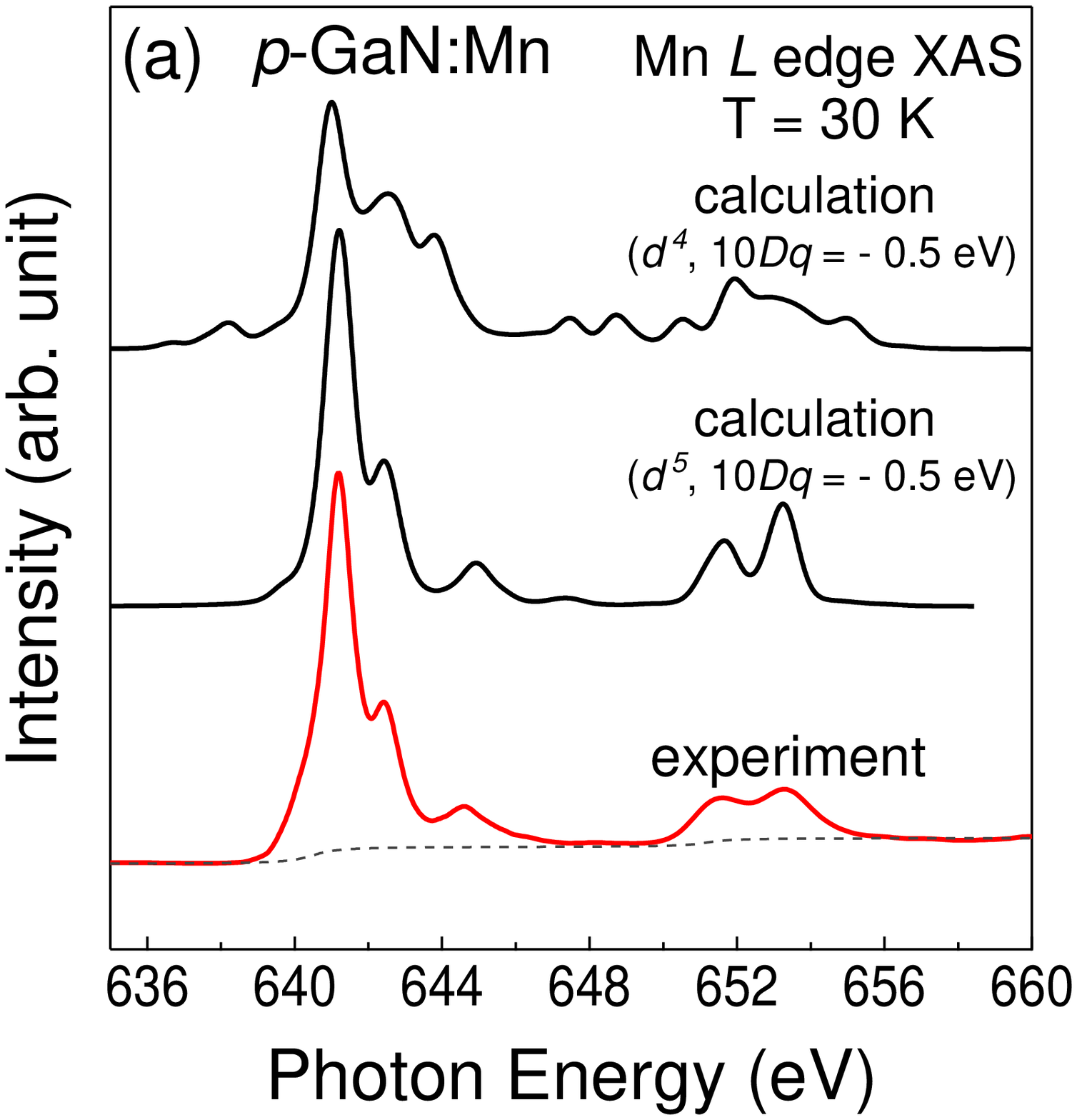}
\includegraphics[height=5cm]{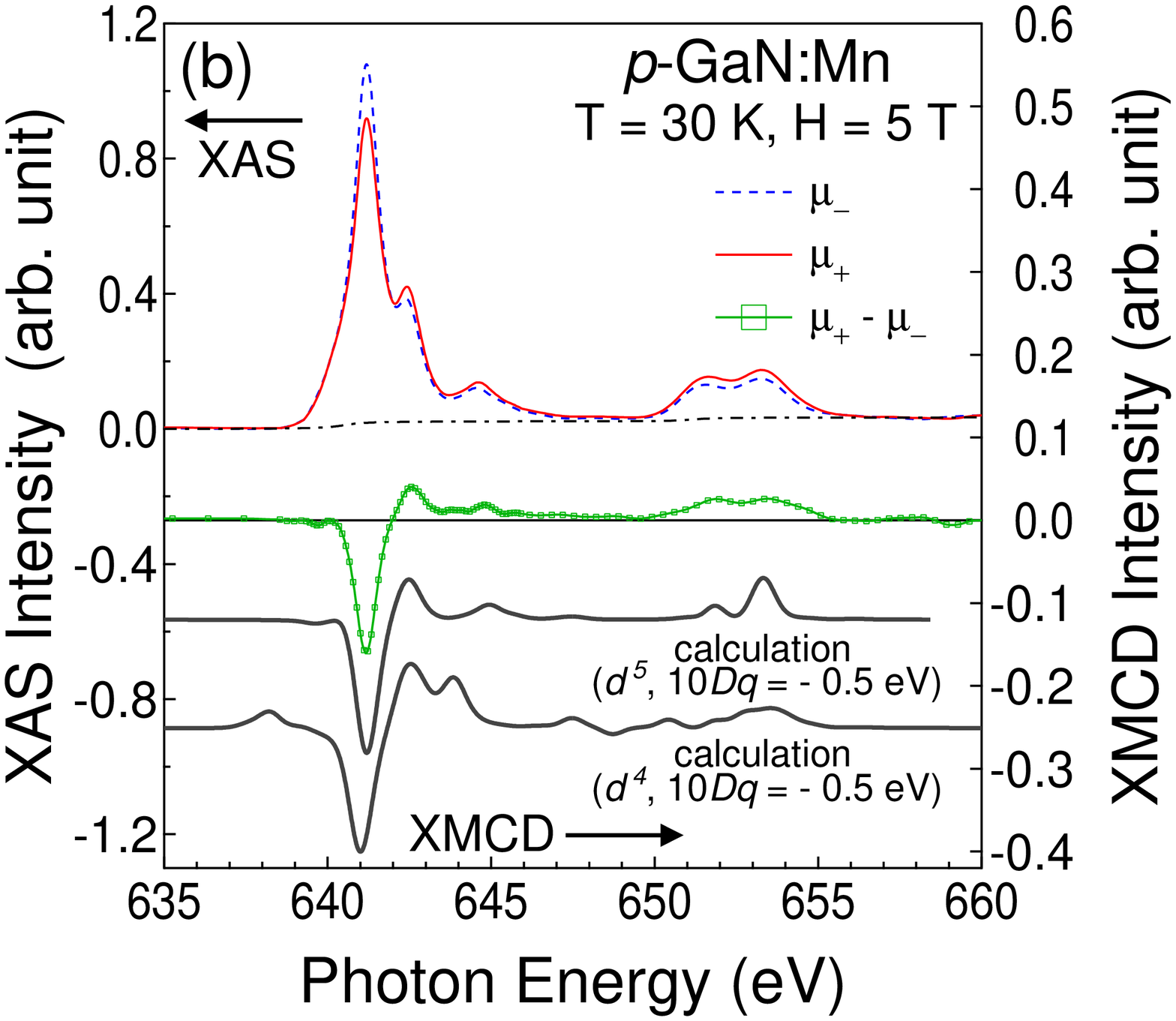}
\includegraphics[height=5cm]{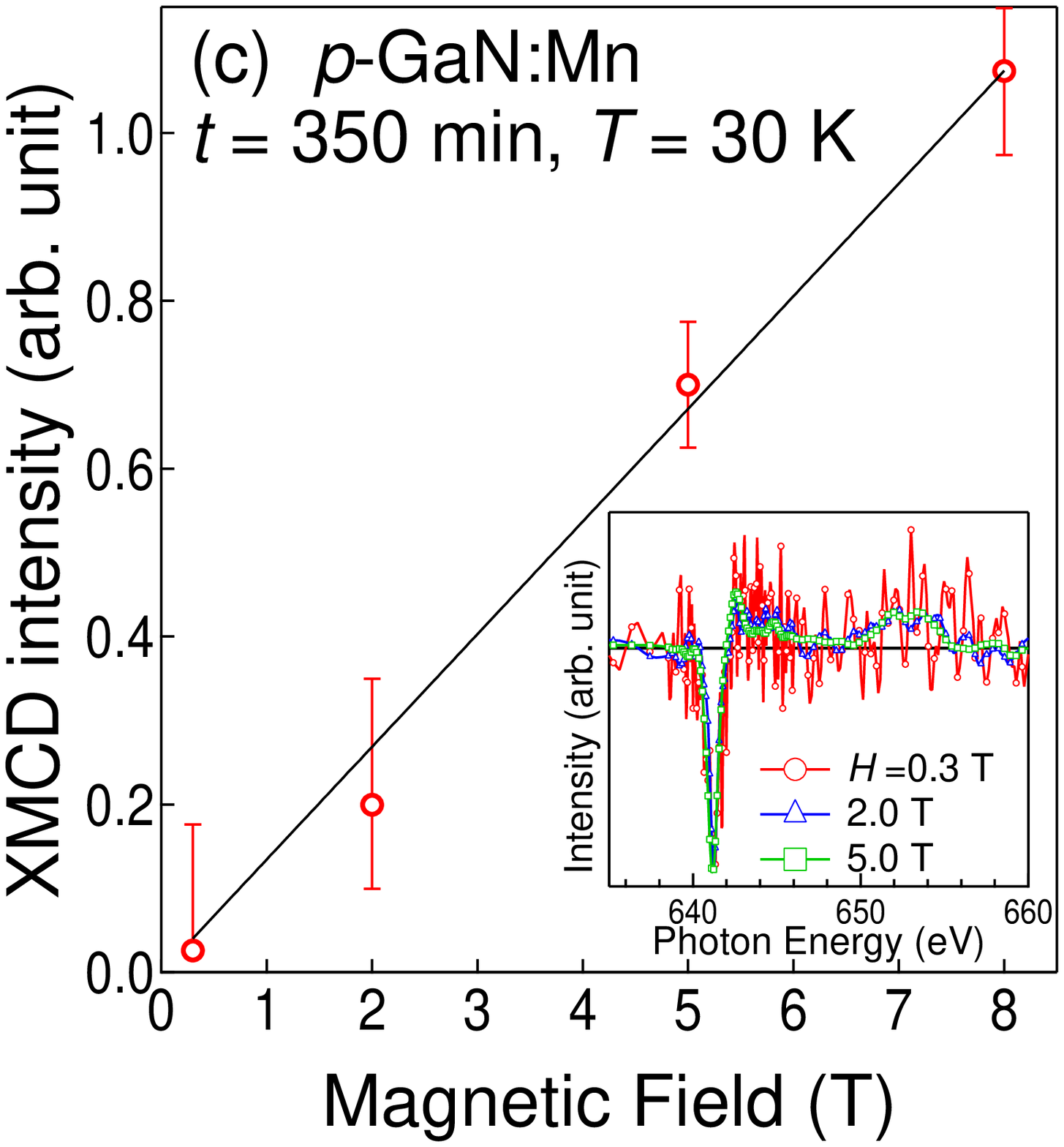}
\label{Mn2pXAS}
\caption{(Color online) Soft x-ray absorption spectra at the Mn $L_{2,3}$ edge of $p$-GaN:Mn. (a) XAS spectra compared with atomic multiplet calculations for the $d^4$ and $d^5$ configuration in the tetrahedral crystal field. (b) XAS and XMCD spectra in a magnetic field of 5 T at 30 K. The atomic multiplet calculations for $d^4$ and $d^5$ are shown in the bottom panels. (c) Magnetic-field dependence of the XMCD signal intensity. The inset shows XMCD spectra recorded at different magnetic field.}
\end{figure}

\vspace{1cm}
\begin{figure}
\includegraphics[height=5cm]{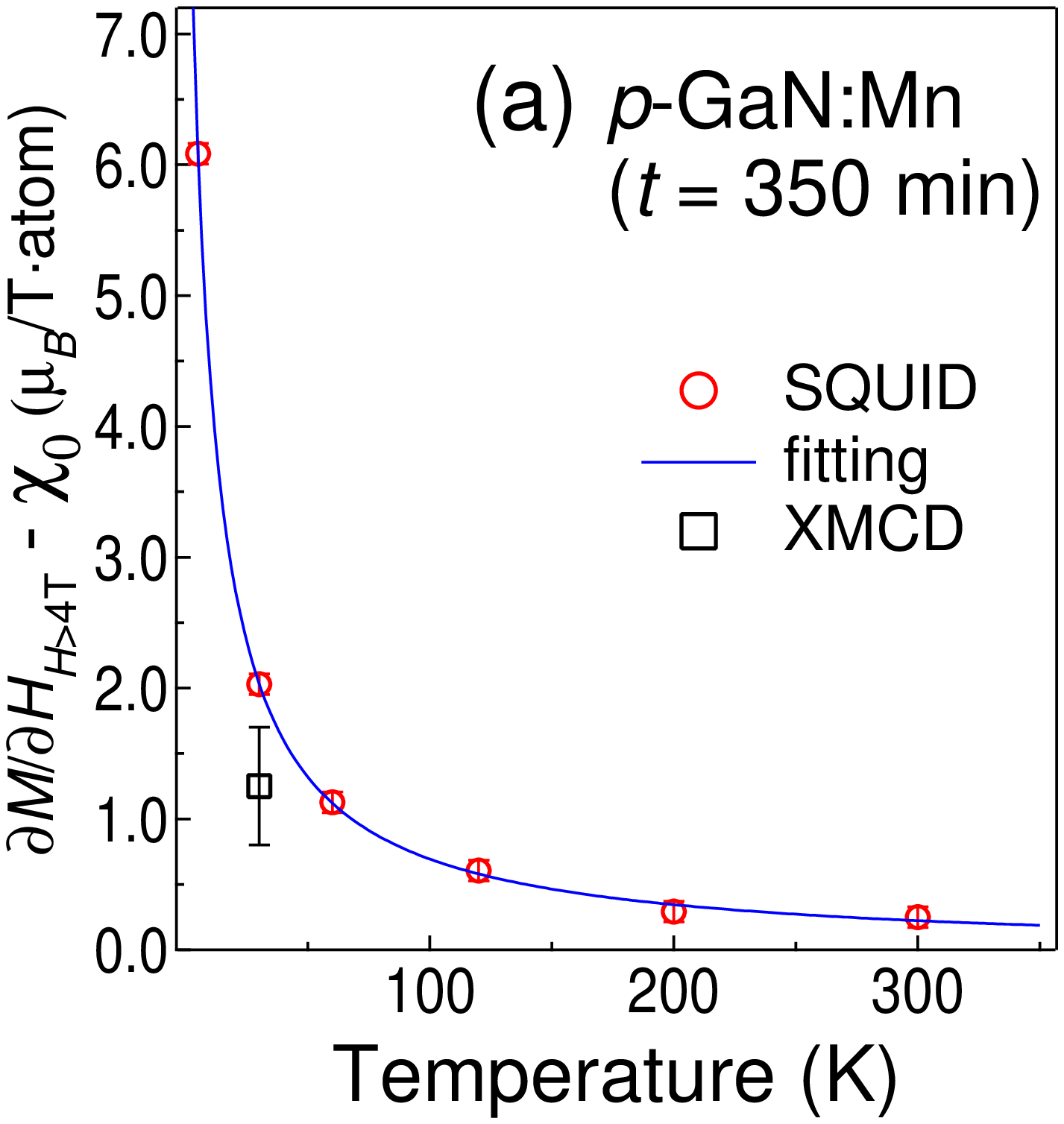}
\includegraphics[height=5cm]{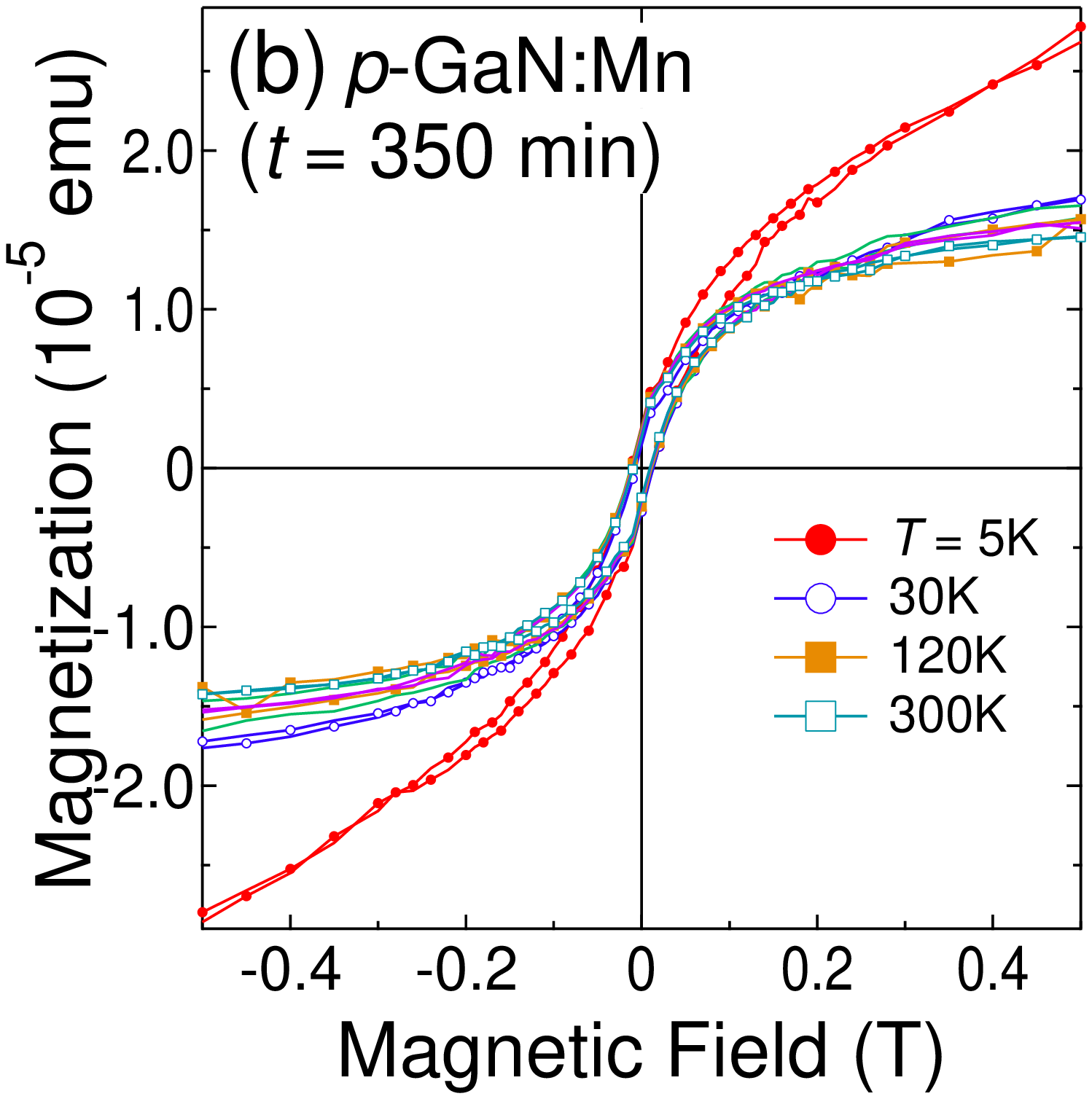}
\includegraphics[height=5cm]{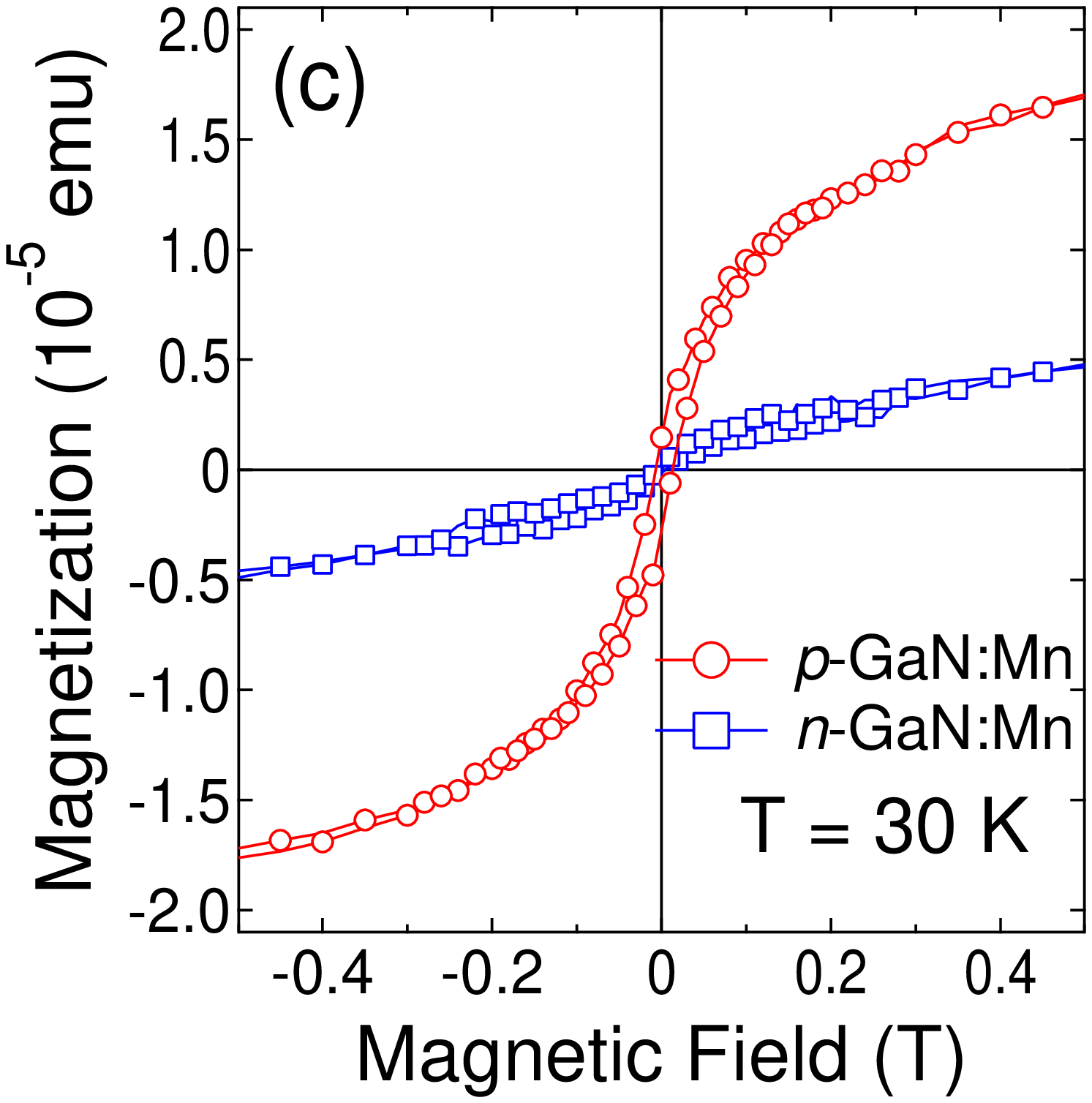}
\label{SQ}
\caption{(Color online) Magnetization curves measured by a SQUID 
magnetometer. (a) Temperature dependence of magnetic susceptibility at high field measured by SQUID. That estimated from XMCD at 30 K is also shown as indicated by square. (b) Magnetization curves of $p$-GaN:Mn at various temperature. (c) Magnetization curves at 30 K of $p$- and $n$-GaN:Mn.}
\end{figure}

\end{document}